\pdfoutput=1
\documentclass[useamsfonts]{pasj01}

\usepackage{graphicx}
\usepackage{graphicx,todonotes}
\usepackage{latexsym}
\usepackage{morefloats}
\usepackage{natbib}
\usepackage{amsmath}
\usepackage{url}
\graphicspath{{figures/}}

\newcommand{\beq}{\begin{equation}}
\newcommand{\eeq}{\end{equation}}

\def\gcm3{\mathrm{g} / \mathrm{cm}^3}

\def\gtsima{$\; \buildrel > \over \sim \;$}
\def\ltsima{$\; \buildrel < \over \sim \;$}
\def\prosima{$\; \buildrel \propto \over \sim \;$}
\def\gsim{\lower.7ex\hbox{\gtsima}}
\def\lsim{\lower.7ex\hbox{\ltsima}}
\def\simgt{\lower.7ex\hbox{\gtsima}}
\def\simlt{\lower.7ex\hbox{\ltsima}}
\def\simpr{\lower.7ex\hbox{\prosima}}

\usepackage{etoolbox}
\makeatletter
\patchcmd{\NAT@citex}
  {\@citea\NAT@hyper@{\NAT@nmfmt{\NAT@nm}\NAT@date}}
  {\@citea\NAT@nmfmt{\NAT@nm}\NAT@hyper@{\NAT@date}}
  {}
  {}
\patchcmd{\NAT@citex}
  {\@citea\NAT@hyper@{%
     \NAT@nmfmt{\NAT@nm}%
     \hyper@natlinkbreak{\NAT@aysep\NAT@spacechar}{\@citeb\@extra@b@citeb}%
     \NAT@date}}
  {\@citea\NAT@nmfmt{\NAT@nm}%
   \NAT@aysep\NAT@spacechar%
   \NAT@hyper@{\NAT@date}}
  {}
  {}
\patchcmd{\NAT@citex}
  {\@citea\NAT@hyper@{%
     \NAT@nmfmt{\NAT@nm}%
     \hyper@natlinkbreak{\NAT@spacechar\NAT@@open\if*#1*\else#1\NAT@spacechar\fi}%
       {\@citeb\@extra@b@citeb}%
     \NAT@date}}
  {\@citea\NAT@nmfmt{\NAT@nm}%
   \NAT@spacechar\NAT@@open\if*#1*\else#1\NAT@spacechar\fi%
   \NAT@hyper@{\NAT@date}}
  {}
  {}
\makeatother

\newcommand{\mtrv}[1]{{\textcolor{red}{#1}}}

\begin{document}
\SetRunningHead{Aihara et al.}{HSC-SSP Survey}

\title{The Hyper Suprime-Cam SSP Survey: Overview and Survey Design}
\author{
Hiroaki~Aihara\altaffilmark{1},
Nobuo~Arimoto\altaffilmark{2,3},
Robert~Armstrong\altaffilmark{4},
St\'ephane~Arnouts\altaffilmark{5},
Neta~A.~Bahcall\altaffilmark{4},
Steven~Bickerton\altaffilmark{6},
James~Bosch\altaffilmark{4},
Kevin~Bundy\altaffilmark{7,8},
Peter~L.~Capak\altaffilmark{9},
James~H.~H.~Chan\altaffilmark{10,11}
% ...
}

\altaffiltext{1}{Department of Physics, University of Tokyo, Tokyo 113-0033, Japan}
\altaffiltext{2}{Department of Astronomy, School of Science, Graduate University for Advanced Studies (SOKENDAI), 2-21-1, Osawa, Mitaka, Tokyo 181-8588, Japan}
\altaffiltext{3}{Subaru Telescope, National Astronomical Observatory of Japan, 650 N Aohoku Pl, Hilo, HI 96720}
\altaffiltext{4}{Department of Astrophysical Sciences, Princeton University, 4 Ivy Lane, Princeton, NJ 08544}
\altaffiltext{5}{CNRS, Laboratoire d'Astrophysique de Marseille, UMR 7326, Aix Marseille Université, F-13388, Marseille, France}
\altaffiltext{6}{Orbital Insight, 100 W. Evelyn Ave. Mountain View, CA 94041}
\altaffiltext{7}{Kavli Institute for the Physics and Mathematics of the
Universe (Kavli IPMU, WPI), UTIAS, University of Tokyo, Chiba 277-8583, Japan}
\altaffiltext{8}{Department of Astronomy and Astrophysics, University of California, Santa Cruz, 1156 High Street, Santa Cruz, CA 95064 USA}
\altaffiltext{9}{California Institute of Technology, 1200 E. California Blvd., Pasadena, CA, 91125, USA}
\altaffiltext{10}{Academia Sinica Institute of Astronomy and Astrophysics, P.O. Box 23-141, Taipei 10617, Taiwan}
\altaffiltext{11}{Department of Physics, National Taiwan University, 10617 Taipei, Taiwan}
% ...
\email{masahiro.takada@ipmu.jp, strauss@astro.princeton.edu}

\maketitle

\begin{abstract}
Hyper Suprime-Cam (HSC) is a wide-field imaging camera on the prime
focus of the 8.2m Subaru telescope on the summit of Maunakea in Hawaii.  A team of
scientists from Japan, Taiwan and Princeton University is using HSC to
carry out a 300-night multi-band imaging survey of the high-latitude
sky.  The survey includes three layers: the Wide layer will cover 1400
deg$^2$ in five broad bands ($grizy$), with a $5\,\sigma$ point-source
depth of $r \approx 26$.  The Deep layer covers a total of 26~deg$^2$ in
four fields, going roughly a magnitude fainter, while the UltraDeep
layer goes almost a magnitude fainter still in two pointings of HSC (a
total of 3.5 deg$^2$).  Here we describe the instrument, the science
goals of the survey, and the survey strategy and data processing. This
paper serves as an introduction to a special issue of the Publications
of the Astronomical Society of Japan, which includes a large number of
technical and scientific papers describing results from the early phases
of this survey.
\end{abstract}

\section{Introduction}
\label{sec:intro}

We live in a golden age for extragalactic astronomy and cosmology.  We
now have a quantitative and highly
predictive
model for the
overall composition and expansion history of the Universe that is in
accord with a large array of independent and complementary
observations.  Observations of galaxies over most of the 13.8 billion
year history of the Universe have led to a broad-brush understanding of
the basics of galaxy evolution.  Studies of the structure of our Milky
Way galaxy are in rough agreement with the current galaxy evolution
paradigm.  However, there are fundamental and
inter-related questions that remain: 
\begin{itemize}
 \item  What is the physical nature of dark matter and dark energy? 
Is dark energy 
 truly necessary, or could the accelerated expansion of the
	Universe be explained by modifications of the law of gravity?
 \item How did galaxies assemble and how did their properties change
       over cosmic time?  Can a coherent galaxy evolution model be
       found that fits both observations of the distant universe, as
       well as detailed studies of nearby galaxies including the
       Milky Way? 
 \item What is the topology and timing of reionization of the
 intergalactic medium at high redshift? What were the sources of
 ultraviolet light responsible for that reionization? 
\end{itemize}

This paper describes a comprehensive deep and wide-angle imaging survey
of the sky designed to address these and other key questions in
astronomy, using the Hyper Suprime-Cam (HSC), a wide-field imaging
camera on the 8.2-meter Subaru telescope, operated by the National
Astronomical Observatory of Japan (NAOJ) on the summit of Maunakea in
Hawaii.  The combination of the large aperture of the Subaru telescope,
the large field of view (1.5 deg diameter) of HSC, and the excellent
image quality of the site and the telescope make this a powerful
instrument for addressing these fundamental questions in modern
cosmology and astronomy.  Under the Subaru Strategic Program (SSP), we
began a survey using both broad- and narrow-band filters in March
2014. The HSC-SSP will use 300 nights of Subaru time over about six
years.  The survey consists of three layers of different solid
angles, going to different depths.  With both the broad- and
narrow-band photometric data, we will explore galaxy evolution over the
full range of cosmic history from the present to redshift 7.  The
measurement of galaxy shapes in the broad-band images will map the
large-scale distribution and evolution of dark matter through weak
gravitational lensing (WL), and allow us to relate it to galaxy
properties and distribution.  Cross-correlations of HSC WL observables
with the spectroscopic galaxy distribution in the Sloan Digital Sky
Survey (SDSS; \citealt{Yorketal:2000})/Baryon Oscillation Spectroscopic
Survey (BOSS; \citealt{Dawsonetal:2013}) and the observed temperature
and polarization fluctuations in the Cosmic Microwave Background (CMB)
will constrain the parameters of the standard model of cosmology, and
test for exotic variations such as deviations from the predictions of
General Relativity on cosmological scales (see \citealt{Weinberg:2013}
for a review).  Studies of the highest-redshift galaxies and quasars
discovered in this survey will lead to a deeper understanding of
reionization, a key event in the thermal history of the Universe.

The HSC survey follows a long tradition of major imaging surveys
in astronomy.  In the modern era, the Sloan Digital Sky Survey
(SDSS; \citealt{Yorketal:2000}) imaged one third of the celestial
sphere with CCDs in five broad bands ($ugriz$), going to a depth of
$r \approx 22.5$. The next generation of imaging surveys has surpassed
SDSS in various combinations of depth, solid angle coverage, and image
quality. For example, the Pan-STARRS1 survey \citep{Chambersetal:2016}
used a 1.8-meter telescope to cover three-quarters of the sky in
$grizy$ almost a magnitude fainter than SDSS.
DECaLS \citep{DECaLS:2016} is covering 14,000 deg$^2$ in $grz$, going
somewhat deeper than Pan-STARRS, and is designed to support the Dark
Energy Spectroscopic Instrument (DESI, \citealt{DESI:2016}).  The Dark Energy
Survey \citep{DES2016} is imaging 5,000 deg$^2$ of the southern
sky
in five bands with the Blanco 4-meter telescope, going to $r \approx
24.3$ ($10\,\sigma$).  Weak lensing cosmology is a key driver of DES,
and similarly is a driver of many of the more recent surveys.  For
example, the CFHT Lens Survey \citep{CFHTLens:2012} covers 154 deg$^2$
in five bands, to $i'=25.5$.  The Kilo-Degree Survey
(KiDS; \citealt{KIDS:2017}) is covering 1,500 deg$^2$ in 4 bands to $r
= 24.9$.  The HSC survey described in this paper goes deeper than all
these surveys, while still covering well over 1,000 deg$^2$, and
including a narrow-band imaging component as well.

This is the first paper in a series describing the HSC survey and
its science in a special issue of the Publications of the Astronomical
Society of Japan.  Other key papers in this issue include a
technical description of the HSC instrument
itself (Miyazaki et al.~2017)
and the software pipelines that analyze
the data (\citealt{Bosch2017,Huang2017}, Murata et al.~2017).
The first year of the data covering over 100 deg$^2$ in five broad
bands have been released to the public, including fully reduced and
calibrated images as well as catalogs of detected objects. The data release is
described in \citet{HSCDR1} (hereafter the HSC DR1 paper).  A separate analysis and catalog of galaxy shapes,
crucial for weak lensing analysis, is included
in \citet{Mandelbaum2017}. 
This special issue also includes more than two dozen science papers based on the early data
from the HSC survey, on topics ranging from asteroids to dwarf
companions of the Milky Way, to weak lensing measurements of clusters,
to some of the highest redshift quasars known.  

We summarize the characteristics of the HSC instrument itself
in \S~\ref{sec:instrument}.  The survey design is described
in \S~\ref{sec:design}, and the observing strategy follows
in \S~\ref{sec:strategy} and \S~\ref{sec:pointing}.  \S~\ref{sec:pipeline} gives a brief overview of
the data processing. We summarize, with a view to the future,
in \S~\ref{sec:conclusion}. 

\section{Hyper Suprime-Cam}
\label{sec:instrument}

\begin{figure}
 \centering{
\noindent \includegraphics[width=0.6\textwidth]{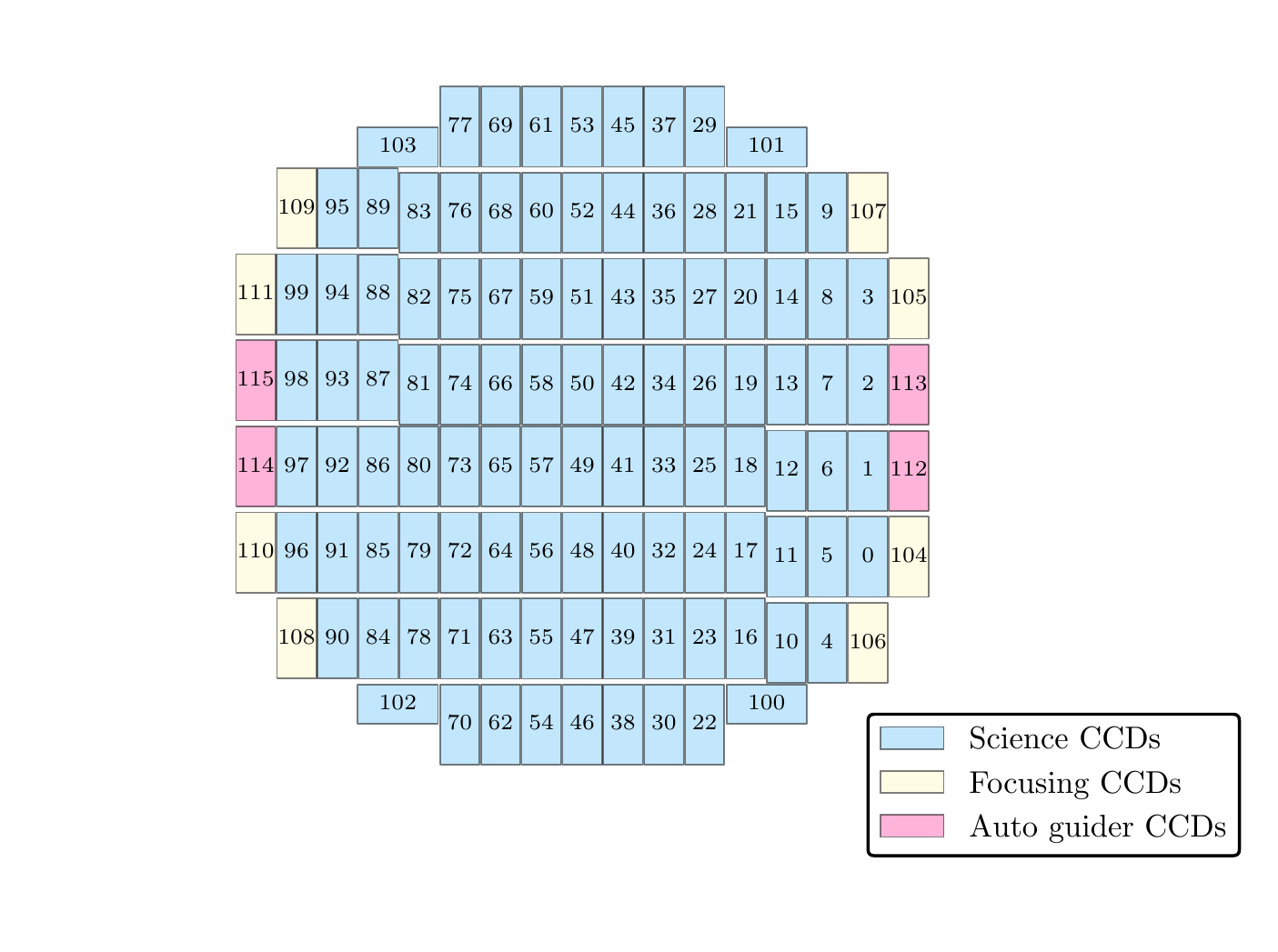}
 \caption{The layout of 116 CCD chips, each of which has 2048$\times
 4096$ pixels ($5.73^\prime \times 11.47^\prime$), 
 on the focal plane. The CCD chips are 
 arranged with two different gaps of approximately 12$^{\prime\prime}$ and 
 $53^{\prime\prime}$ between the neighboring chips. The focal plane is
 approximately $1.5^\circ$ in diameter. There are 104 science chips
 (indicated in blue),  4 chips used for auto-guiding (in yellow) and
8 chips for monitoring  
the focus (in light red). Each chip is identified with a number from 0
to 115.
 \hfill
 \label{fig:ccdlayout}}}
\end{figure}
\begin{figure}
 \centering{
 \includegraphics[scale=0.42]{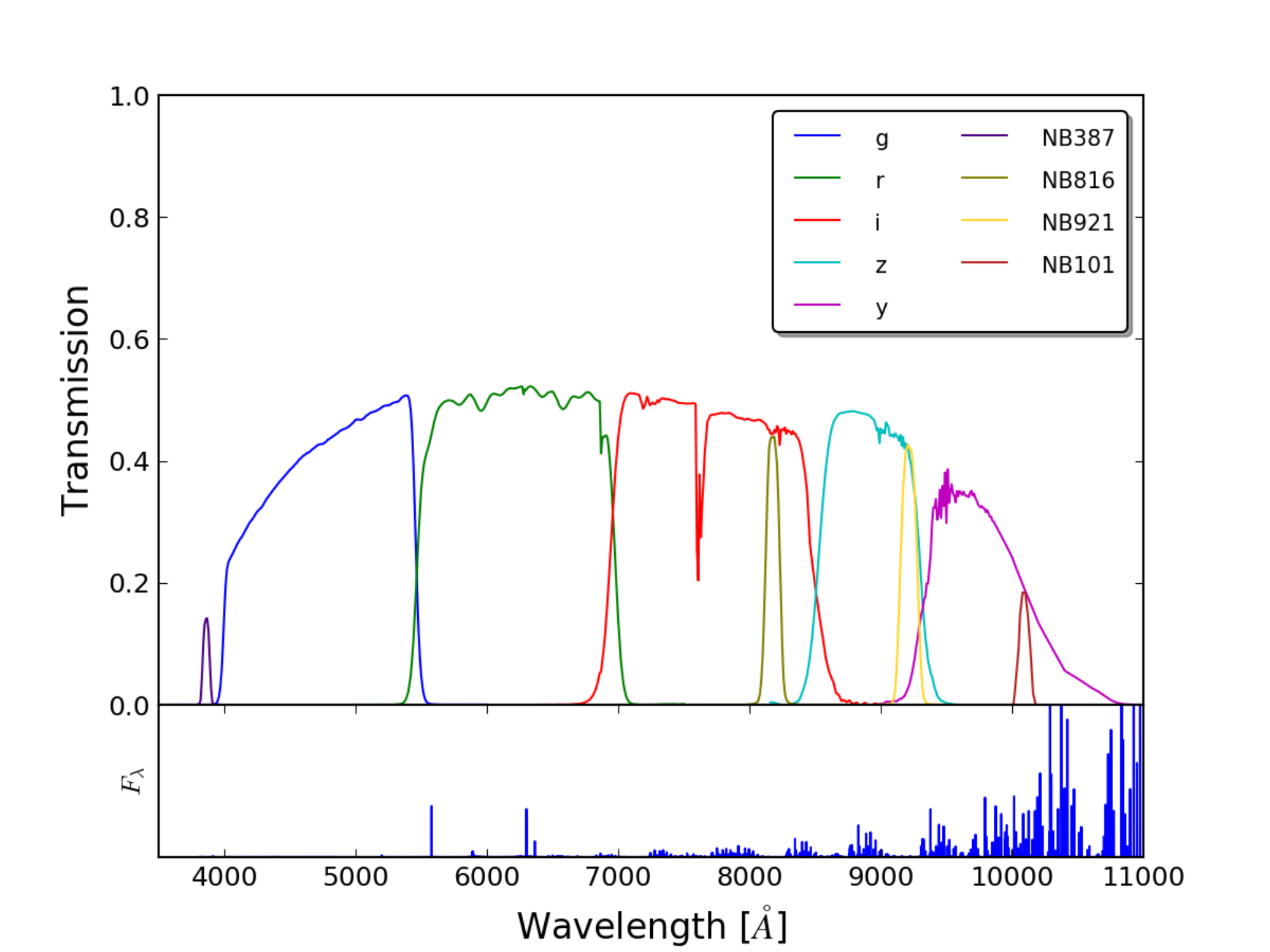}
 \caption{The HSC bandpasses, including the reflectivity of all mirrors,
 transmission of all optics and filters as well as the atmosphere, and
 response of the CCDs, 
 assuming an airmass of 1.2. Both the broad-band and narrow-band filters
 used in the survey are shown. The lower panel shows the spectrum of
 sky emission lines, demonstrating that the red narrow-band filters
 lie in relatively dark regions of the sky spectrum. \hfill
\label{fig:filters}
 }
}
\end{figure}
While there are 
other
8-meter class telescopes around the world, Subaru
is the one with by far the largest field of view.  Suprime-Cam
\citep{Miyazakietal:2002}, with its $\sim 0.25 \,\rm deg^2$ field of
view and superb delivered image quality (routinely $0.6''$ FWHM), has
been a world leader in wide-field studies of the distant and faint
Universe \citep[e.g.,][]{Iyeetal:2006,Furusawaetal:2008}.  Hyper Suprime-Cam (HSC), its
successor, takes advantage of the full 
accessible field of view of the Subaru telescope (1.5$^\circ$ diameter),
and thus has a survey power about 8 times larger than that of
Suprime-Cam.  The speed with which a given facility can survey
the sky to a given depth is proportional to the product of the collecting
area of the telescope and the field-of-view of the camera
(\'etendue), although it also depends on image quality and the
fraction of time any given facility is devoted to survey work.  
The \'etendue of HSC is the
largest of all existing wide-field optical imaging cameras, not to be
surpassed until the Large Synoptic Survey Telescope
(LSST; \citealt{LSSTScienceBook,LSSTOverview}) sees first light
in late 2019.

\begin{table*}
\begin{center}
\caption{{Hyper Suprime-Cam Characteristics\hfill}
\label{tab:instrument}
}
\begin{tabular}{p{2.5in}p{3.5in}}
\hline
\hline
Instrument weight & $\sim$3.2 tons\\
Field of View & 1.5$^\circ$ diameter\\
Vignetting & 0 at $0.15^\circ$; 26\% at edge\\
Pixel scale & 15$\mu\rm m \simeq 0.168''$\\
Detectors & Hamamatsu Photonics KK 2048x4096 (S10892-02)\\
Number of CCDs & 104 science, 4 autoguide, 8 for monitoring the focus$^{(1)}$\\
CCD QE &40\% at 4000\AA, 10,000\AA; 95\% at peak (at -100${}^{\circ}$C)\\
CTE & 0.999999\\
Gain & 3.0$e^{-1}$/ADU \\
 Readnoise & 4.5 e$^-$\\
 Readout & 20 sec$^{(2)}$ \\
 Saturation level & 150,000e$^-$\\
Filters & $grizy$ + 4 narrow-band$^{(3)}$ \\
Filter Exchanger & 6 filters installed at a time\\
Filter Exchange Time & $\sim $30 minutes$^{(4)}$ \\ \hline
\end{tabular}
 \end{center}
{\footnotesize NOTES -- (1) The camera has 104 science CCDs, 4 CCDs for
auto-guiding, and 8 CCDs for monitoring the focus (also see
\url{http://subarutelescope.org/Observing/Instruments/HSC/parameters.html}).
(2) 20 sec is the time needed for reading out signals from CCDs. The
actual overhead time, the time from the end of the previous exposure to
the beginning of the next exposure, is about 35 sec, which includes time
needed for transferring data from the instrument to the data-taking
computers, slewing the telescope between dithering positions, and so
on. (3) The survey described in this paper uses five broad-band and four
narrow-band filters, but there are additional filters available for use
with HSC.  (4) Before exchanging filters in the filter exchange unit,
one must first move the telescope to zenith, rotate the instrument to a
fiducial angle, and close the primary mirror cover. \hfill}
\end{table*}
\begin{table*}
{
\begin{center}
\begin{tabular}{l|llllll}
\hline\hline
Layer & Area 
& \# of 
& Filters \& Depth 
& Volume
& Key Science \\
& [deg$^2$] 
&pointings
&
& [$h^{-3}$Gpc$^3$]
 \\
\hline
Wide& 1400 &916 & $grizy~ (i\simeq 26)$ 
& $\sim 4.4 (z<1.5)$
&WL Cosmology, $z\sim 1$ gals, Clusters\\
Deep& 26 &15& $grizy$+3NBs ($i\simeq 27$)
& $\sim 0.5 (1<z<5)$
&$z\simlt 2$ gals, SNeIa, WL calib.
\\
UltraDeep& 3.5 & 2
&$grizy$+3NBs ($i\simeq 28$)
& $\sim 0.07 (2<z<7)$
& 
high-$z$ gals (LAEs, LBGs), SNeIa
\\
\hline
\end{tabular}
\caption{Summary of HSC-Wide, Deep and UltraDeep layers. \hfill\label{tab:design}
}
\end{center}
}
\end{table*}
The HSC instrument (\citealt{Miyazakietal:2015}, Miyazaki et al.~2017) was designed and built by
an international team involving scientists from NAOJ, the Kavli
Institute for the Physics and Mathematics of the Universe (Kavli IPMU),
the University of Tokyo, the Academic Sinica Institute of Astronomy and
Astrophysics in Taiwan (ASIAA), and Princeton University.  The
instrument has a large and optically very sophisticated seven-element
Wide-Field Corrector (WFC), designed and built by Canon, Inc.  The WFC
incorporates an Atmospheric Dispersion Corrector (ADC) and delivers an
instrumental Point-Spread Function (PSF) for which the diameter
enclosing 80\% of the light ($D_{80}$) is $0.2''$ or better over the
entire field in all filters.  A Prime Focus Unit (PFU) built by
Mitsubishi Electric Corporation, which incorporates a precise hexapod
for attitude and focus adjustment, holds the WFC and the camera in place
at the telescope prime focus. The entire structure is roughly 3 meters
tall, and weighs almost 3 tons. The corrector gives an unvignetted field
of view to a diameter of 10 arcmin; vignetting is a roughly linear
function of field radius from that point, reaching 26\% at the edge of
the field at $0.75^\circ$.  The Subaru top-end structure has been
modified to accommodate the PFU and WFC.  The WFC can be used by other
wide-field instruments as well, and is incorporated into the design of
the planned Prime Focus Spectrograph \citep{Takadaetal:2014,TamuraPFS}.

Table~\ref{tab:instrument} summarizes the characteristics of the HSC
instrument.  The focal plane is paved with a total of 116 Hamamatsu Deep
Depletion CCDs, each $2\,\rm K \times 4K$ pixels. 
The layout of the CCD chips is shown in
Figure~\ref{fig:ccdlayout}
(see \url{https://www.subarutelescope.org/Observing/Instruments/HSC/ccd.html} 
for details).
The 15$\mu$m pixels
subtend about $0.168''$ on the sky, with some modest variation over the
focal plane.  Four of the CCDs are used for guiding and eight for
automatically monitoring focus, leaving 104 science detectors.  These
chips are three-side buttable and each have four independent readout
amplifiers.  Gaps between chips are small, typically $12''$ in one
dimension and $53''$ in the other.  Regions of the focal plane more than
49 arcmin from the center are masked, giving an effective area of about
1.77~deg$^2$. The chips have excellent characteristics: low read noise,
excellent charge transfer efficiency, few cosmetic defects, and most
importantly, high quantum efficiency ($>40\%$) from 4000\AA\ to
10,000\AA\ (blueward of 4000\AA, the response is limited both by the
CCDs and the optical elements in the WFC).  The model system response,
including reflectivity and transmission of all optics, is shown in
Figure~\ref{fig:filters}.  Kawanomoto et al.~(2017) give a detailed
description of the filters used in the HSC SSP.  With $0.168''$ pixels,
the images are well-sampled in even the best seeing seen on the
instrument, $0.4''$.

  Table~\ref{tab:filters} lists the characteristics of the filters
  used in the HSC-SSP survey. 
 The effective wavelength is defined following \citet{Schneider:1983}
 as
\begin{equation} 
\lambda_{\rm eff} = \exp\left[{\int d \ln \lambda\, S(\lambda) \ln \lambda} \over
{\int d \ln \lambda\, S(\lambda)}\right],
\label{eq:lambda_eff}
\end{equation}
where $S(\lambda)$ is the throughput of the telescope, camera, filter
and atmosphere as a function of wavelength (as plotted in
Figure~\ref{fig:filters}).  We characterize the width of the filters
in two ways.  The full-width at half-maximum (FWHM) is a standard
measure.  Alternatively, we can follow \citet{Schneider:1983} to
define the dimensionless quantity:
\begin{equation} 
\sigma^2 = {{\int d \ln \lambda\, S(\lambda) \ln^2(\lambda/\lambda_{\rm eff})} \over
{\int d \ln \lambda\, S(\lambda)}}.
\label{eq:sigma}
\end{equation}
In the limit of a narrow top-hat filter: 
\begin{equation} 
{\rm FWHM} = \lambda_{\rm eff}\left(e^{\sqrt{3}\sigma} -
e^{-\sqrt{3}\sigma}\right).
\end{equation}
We also tabulate the effective filter throughput $Q$, defined as the
integral of $S(\lambda)$ over $\ln \lambda$ (i.e., the denominator of
equations \ref{eq:lambda_eff} and \ref{eq:sigma}).
\begin{table*}
\begin{center}
\begin{tabular}{lllll}
\hline
\hline
Filter & $\lambda_{\rm eff}$ (\AA) & $\sigma$ & FWHM (\AA) & $Q$\\ \hline
$g$ & 4754 & 0.0873 &  1395 & 0.1590 \\
$r2$ & 6175 & 0.0711  &  1503 & 0.1667 \\
$i2$ & 7711 & 0.0599 &  1574 & 0.1149 \\
$z$ & 8898 & 0.0263 &   766 & 0.0487 \\
$y$ & 9762 & 0.0316 &  783 & 0.0352 \\
$r$ (old) & 6178 & 0.0694 &   1495 & 0.1505 \\
$i$ (old) & 7659 & 0.0597 &   1523 & 0.1154 \\
NB$387$ & 3863 & 0.0049 &   55 & 0.0026 \\
NB$816$ & 8177 & 0.0049 &   113 & 0.0074\\
NB$921$ & 9214 & 0.0050 &   134 & 0.0075\\
NB$101$ & 10092 &  0.0039 &  91 & 0.0018 \\
\hline \hline
\end{tabular}
\caption{HSC Filter characteristics.  For each filter, we include the
effective wavelength, a fractional width defined in
equation~(\ref{eq:sigma}), the FWHM, and a measure of
total throughput, as defined in the text.  Note that we include the characteristics both of the
updated $r$ and $i$ filters ($r2$ and $i2$) as well as the old
versions used in the early part of the survey.
\label{tab:filters}\hfill}
\end{center}
\end{table*}

The instrument is installed at prime focus using the existing Top Unit
Exchanger (TUE) instrument handler, though modifications were necessary
to ensure that the instrument is not damaged during installation.

The camera has a roll-type shutter, with excellent timing accuracy,
allowing uniform exposure time over the field of view.  Including
readout and all overheads, the minimum time between exposures is about 35
seconds, allowing for efficient surveying of the
sky. The overhead for telescope slewing is negligible, because we can
move the telescope to the next target field during the camera readout.
The filter exchange mechanism can hold six filters at
one time.  With the telescope at zenith, changing filters takes about
10 minutes.  However, for safety reasons the primary mirror cover
needs to be closed and instrument rotated to a fiducial angle before
the filters can be changed, meaning that it takes about 30 minutes in
practice between the end of one sky exposure and the start of the
following exposure in a different filter.  

The HSC survey uses five broad-band filters ($grizy$) modeled on the
SDSS filter set (Figure~\ref{fig:filters} and Table~\ref{tab:filters}), as
well as four narrow-band filters sensitive to emission lines such as 
the Lyman-$\alpha$ line over a wide range of redshifts.
Given the filter exchange time, we usually do no more
than two filter exchanges per night in survey mode.

HSC saw astronomical first light in 2012 August, and became a general
user instrument for Subaru in March 2014, when the survey described here
began.  The original $r$ and $i$-band filters installed in the camera
did not meet our full specifications, and we replaced them with
filters with significantly more uniform response across the focal
plane.  
We have used the new $i$-band filter (HSC-$i2$) since
Feb 2016 and the new $r$-band filter (HSC-$r2$) since Jul
2016\footnote{\url{http://subarutelescope.org/Observing/Instruments/HSC/sensitivity.html}}.
The DR1 data 
were all taken with the older versions of the
filters.

\section{Survey design}
\label{sec:design}

\begin{table*}
  \begin{center}
    \begin{tabular}{l|ccccccccc}
      \hline\hline
      {\bf Wide}            &  $g$  &  $r$  &  $i$  &  $z$   &  $y$  &          &          &           & \\
      total exp.~(min)        &  10  &  10   &   20
		 &   20   &   20  &          &          &           & \\
      {exp. per visit (min/sec)} & 2.5 & 2.5 & 200~sec & 200~sec & 200~sec\\
      depth (mag)    & 26.5  &  26.1 &  25.9 &  25.1  &  24.4 &          &          &           & \\
      \hline \hline
      {\bf Deep}            &  $g$  &  $r$  &  $i$  &  $z$   &  $y$  &  NB387 & NB816  &  NB921 & \\
      exp. (min)        &  84   &  84   &   126  &   210   &   126
			     &  84      &  168     &
					 252\\
      {exp. per visit (min)} & 3 & 3 & 4.5 & 4.5 & 4.5 & 14 & 15 & 15\\
      depth (mag)    & 27.5  &  27.1 &  26.8 &  26.3  &  25.3 &  24.5    & 25.8     &  25.6    & \\
      \hline \hline
     {\bf  UltraDeep}       &  $g$  &  $r$  &  $i$  &  $z$   &  $y$  &          & NB816  &  NB921 & NB101\\
      exp. (min)        &  420   &  420   &  840   &
		     1134   &  1134   &   ---       & 630 
				 &  840      & 1050\\
     {exp. per visit (min)} & 5 & 5 & 5 & 5 & 5 & --- & 15 & 15 & 15\\
      depth (mag)    & 28.1  &  27.7 &  27.4 &  26.8  &  26.3 &  ---        & 26.5    &  26.2    & 24.8\\
       \hline \hline
    \end{tabular}
    \caption{ Total exposure time (min), exposure time of each visit,
and target depth ($5\sigma$ depth for a point source) for each filter in
the Wide, Deep and UltraDeep layers, respectively.  Dividing the total
exposure time by each visit exposure time gives the number of visits for
each field.  For example, in the Wide layer, there are four visits for
each field in $g$ and $r$, and six visits in $i$, $z$, and $y$.  \hfill
} \label{tab:exptime}
  \end{center}
\end{table*}
The HSC-SSP has been awarded 300 nights. We originally planned to
observe 60 nights a year for five years starting in 2014, but the first
two years of observing were allocated substantially less than this
amount, as the instrument was operated under a shared-risk mode,
available only for a few months each semester, while the Subaru
Observatory was refining its techniques for mounting, operating, and
unmounting this complex instrument.  Thus the survey is likely to
continue into 2019.

As described above, the principal scientific goals of the survey are
studies of the distribution of dark matter in the universe from WL
measurements, and probing the evolution of galaxies.  This motivates
surveying in three different modes, or layers: Wide, Deep, and
UltraDeep, as summarized in Table~\ref{tab:design}.

While the Wide, Deep and UltraDeep layers will all take data in the same
five broad-band filters, $grizy$ (Figure~\ref{fig:filters}), they use
different narrow-band filters, they go to different depths, and they
cover different solid angles and thus cosmological volumes.
Table~\ref{tab:exptime} summarizes the survey parameters of each
layer.  The survey depths are defined as the $5\,\sigma$ PSF
magnitude limit for isolated point sources, using photometric errors
determined by the imaging pipeline (\citealt{Bosch2017}; see also the
discussion in the HSC DR1 paper).
The saturation limits depend on the individual visit exposure times,
the seeing, and the sky brightness and transparency.  For point sources in typical conditions for the exposure times in the Wide layer, the saturation limits in $grizy$ are roughly 
\mtrv{17.8, 17.8, 18.4, 17.4, and 17.1}, 
respectively.  These are for median seeing of $0.72^{\prime\prime}, 0.67^{\prime\prime}, 0.56^{\prime\prime}, 0.63^{\prime\prime}$ and $0.64^{\prime\prime}$, respectively; the saturation limit in seeing a factor $f$ larger will be fainter by $5\,\log f$ magnitudes.  Similarly, the saturation limit in the 30-second exposures (Section 3.3) is brighter by $2.5\,\log g$ magnitudes, where $g$ is the ratio of the single-visit exposure time to 30 seconds (corresponding to 1.75 mags in $gr$ and 2.06 mags in $izy$ in the wide survey).  Finally, the throughput of the system is proportional to the quantity $Q$ tabulated in Table~\ref{tab:filters}, which allows one to calculate approximate saturation limits in the narrow-band filters as well.

\subsection{Survey fields}

\begin{table*}
\begin{center}
 \begin{tabular}{l|l|l} \hline\hline
 Layer & Field name & (RA, Dec) \\ \hline
Wide & 
 North field & $13:20\le {\rm RA}\le 16:40$ \& $42^\circ\le {\rm
	  dec}\le 44.5^\circ$ ($\simeq 90$ deg$^2$) \\ \cline{2-3} 
 & Spring equatorial field & $08:30\le {\rm RA}\le 15:00$ \& $-2^\circ \le
	  {\rm dec}\le +5^\circ$  
	   ($\simeq 680$ deg$^2$)\\ \cline{2-3}
& Fall equatorial field & $22:00\le {\rm RA}\le 02:40$ \&
	  $-1^\circ\le {\rm dec}\le 7^\circ$  ($\simeq 630$ deg$^2$)  \\ \hline 
 Deep & XMM-LSS & ($02:23:15.33,-05:18:30.67$)\\ 
& & ($02:22:22.43,     -04:03:44.48$) \\
& & ($02:27:19.86,     -04:27:47.42$) \\ \cline{2-3}
&E-COSMOS & $(09:57:28.60, +02:57:21.00)$\\
& & $(10:03:28.60,     +02:57:21.00)$\\
& & $(09:57:28.60,     +01:27:21.00)$\\
& & $(10:03:28.60,     +01:27:21.00)$\\ \cline{2-3}
&ELAIS-N1 & $(16:11:00.81,     +53:56:30.71)$\\
& & $(16:15:46.66,     +54:59:17.74)$\\
& & $(16:10:54.07,     +56:02:49.65)$\\
& & $(16:06:16.19,     +54:59:16.91)$\\ \cline{2-3}
& Deep2-3 & $(23:32:08.46,     +00:16:49.43)$\\
& & $(23:27:02.16,     +00:16:49.41)$\\
& & $(23:29:26.53,     -00:50:38.94)$\\
& & $(23:24:12.60,     -00:49:46.61)$ \\ \hline
 UltraDeep  & SXDS & $(02:18:15.60, -04:51:00.00)$ 
\\ \cline{2-3}
& COSMOS & $(10:00:28.60, +02:12:21.00)$ \\ \hline
-- & AEGIS & $(14:17:00.0, +52:30:00.00) $\\ 
\hline\hline
 \end{tabular}
 \caption{List of the target fields; see Figures~\ref{fig:wide_fields}
 and \ref{fig:deep_fields}.  The coordinates for the Deep and UltraDeep
 layers and the AEGIS (All-wavelength Extended Groth Strip;
 \citealt{2007ApJ...660L...1D}) field are the centers of each fiducial
 pointing. The range of coordinates for the Wide layer defines the
 approximate boundaries of each survey field, although the exact
 boundary of the full depth region depends on the details of the dither
 pattern, as described in \S~\ref{sec:pointing}.  The numbers in the
 round brackets for the Wide layer give the area of each field.  AEGIS
 is a single pointing to Wide layer depth in the five broad bands,
 useful for calibrating photometric redshifts.\hfill
 \label{tab:field_names}}
\end{center}
\end{table*}
Table~\ref{tab:field_names} summarizes the target fields for the
HSC-Wide, Deep and UltraDeep layers (also see Figure~\ref{fig:wide_fields}). 
The fields are chosen to overlap the footprint of the Sloan Digital Sky
Survey \citep{Yorketal:2000,SDSSDR8} and the Pan-STARRS1
survey \citep{Chambersetal:2016}, as we use them for the first-order photometric and
astrometric calibration. In the following we describe details of
the target fields and the rationale for selecting these fields for each layer.

\subsubsection{HSC Wide layer}
\label{sub:wide}

One of the primary science drivers for the HSC-Wide layer is to explore
the nature of dark matter and dark energy via WL observables
\citep[e.g.,][]{Weinberg:2013,TakadaJain:04,OguriTakada:11}.  We will
primarily use the $i$-band data to perform the galaxy shape
measurements.  We perform $i$-band observations in the Wide layer when
the weather is clear and the seeing is good, in order to perform
accurate shape measurements of galaxies (see below for details).  At a
depth of $i\approx 26$, we predict a weighted mean number density of
galaxies for which shapes can be measured of $\bar{n}_{\rm eff}\simeq
20$~arcmin$^{-2}$, with a mean redshift of $\langle z\rangle\simeq 1$.
Combining the $i$-band data with the $grzy$ photometry will allow us to
estimate photometric redshifts (photo-$z$) for every galaxy used in the
WL analysis; the relative depths of the different bands are selected to
optimize the photo-$z$ accuracy \citep{Tanaka:2017}.  A solid angle of
1,400~deg$^2$ will give us the statistical precision for the WL
observables to obtain a tight constraint on dark energy parameters at a
similar level of precision to that of Stage-III dark energy experiments
\citep{Mandelbaum2017}.

\begin{figure*}[t]
 \centering{
\noindent \includegraphics[width=0.45\textwidth]{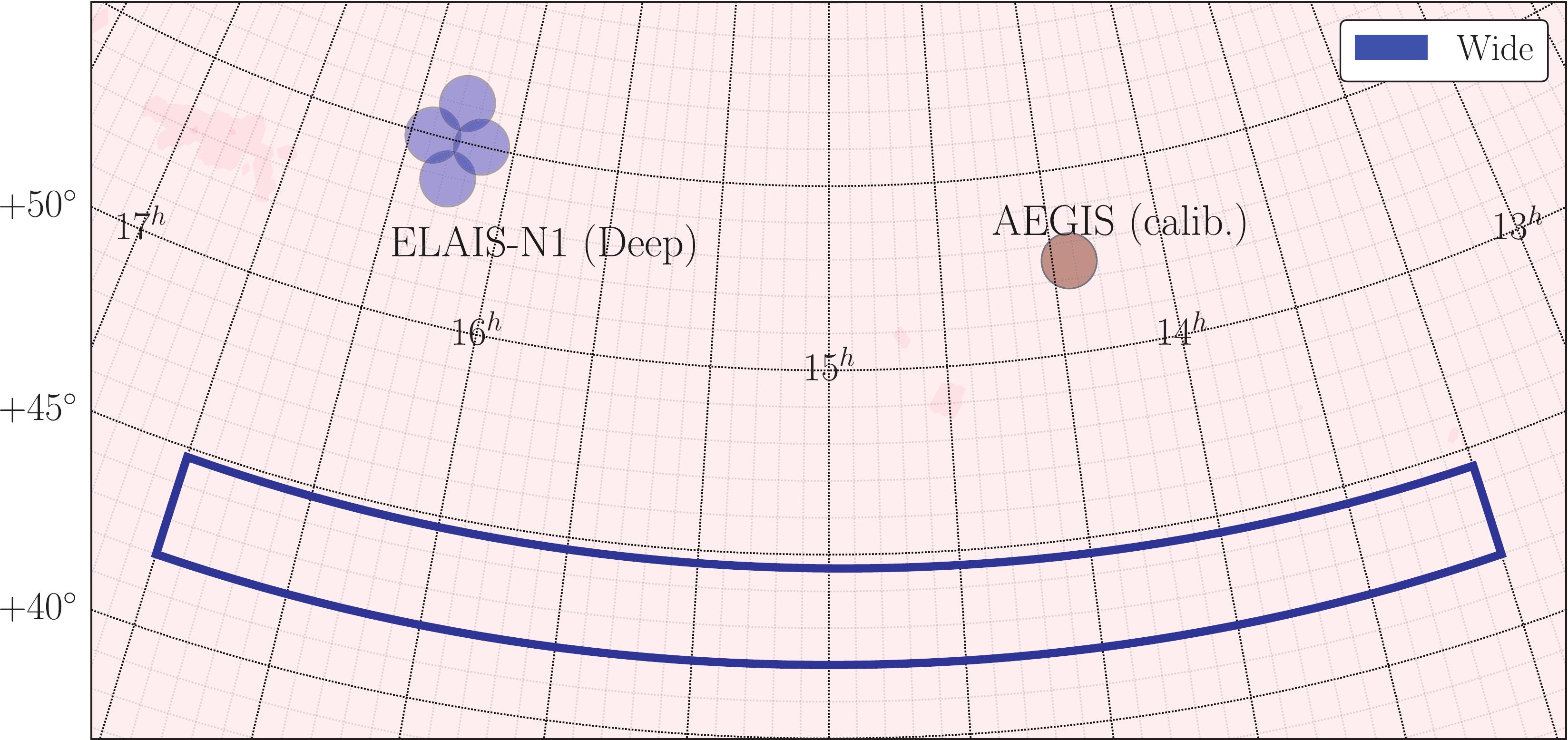}\\
 \vspace*{1em}
\noindent\includegraphics[width=0.95\textwidth]{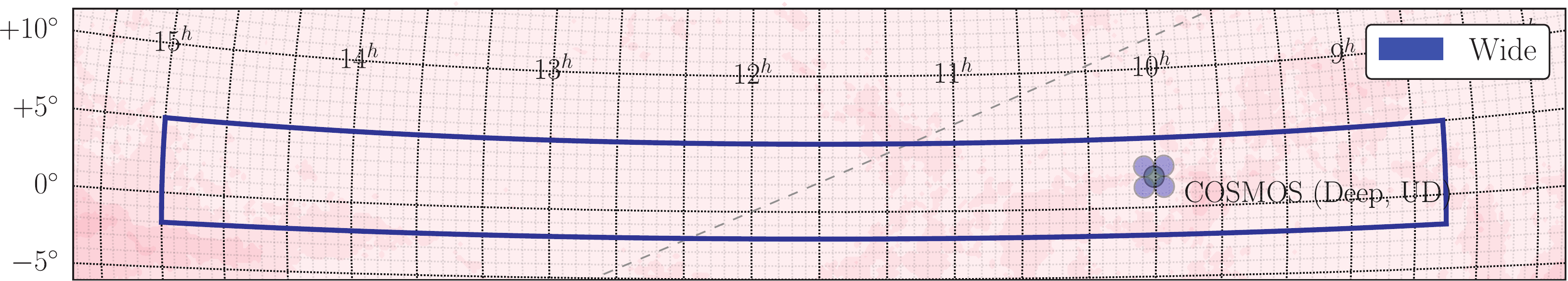}\\
 \vspace*{1em}
\noindent\includegraphics[width=0.9\textwidth]{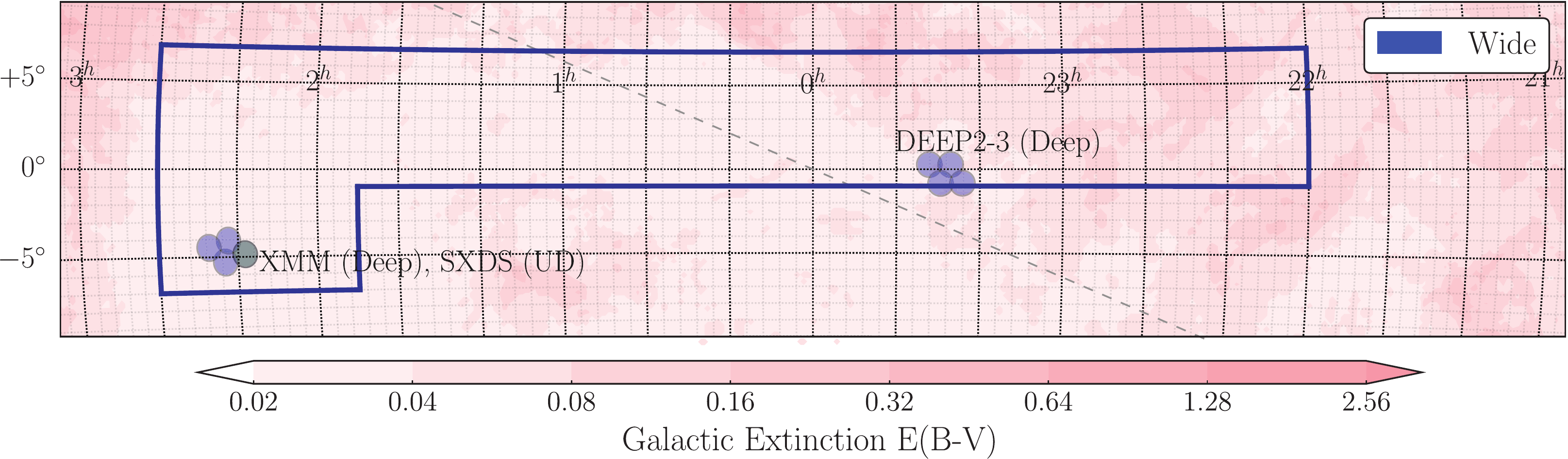}
 }
 \caption{The location of the HSC-Wide, Deep, and UltraDeep fields and
 the AEGIS field 
 on the sky in equatorial coordinates. The color scale gives the level of 
Galactic dust extinction from \citet{SFD}, as denoted by the color
bar.  See Figure~\ref{fig:deep_fields} for the details of the Deep and
UltraDeep fields.\hfill
 \label{fig:wide_fields}}
\end{figure*}
The Wide layer sky coverage is mostly along the Celestial Equator
(making the fields easily observable from both hemispheres).  The
Wide layer fields stretch
over a wide range of right ascension, such that fields are reachable at
all times of the year.  We selected regions of sky low in Galactic
extinction, away from the disk of the Milky Way.  The survey footprint
includes three large spatially contiguous regions, to enable
cosmological analyses on large scales.  Figure~\ref{fig:wide_fields} and
Table~\ref{tab:field_names} show the survey footprint, which consists
of three parts, termed
``the fall equatorial field'', ``the spring equatorial field'' and ``the
north field''.  Our selection of these regions is aimed to
overlap other multi-wavelength data to maximize the scientific synergy
with HSC.  In particular, we consider the arc-minute-resolution,
high-sensitivity CMB survey by the Atacama Cosmology Telescope (ACT;
\citealt{Swetzetal:2011}) in Chile, and its polarization extension
ACTPol \citep{Thorntonetal:2016}; X-ray data from the XMM-XXL survey
\citep{XXL} and eROSITA\footnote{\url{http://www.mpe.mpg.de/eROSITA}};
near-/mid-infrared imaging surveys such as
VIKING\footnote{\url{http://www.astro-wise.org/projects/VIKING/}}  \citep{VIKING2013}
and
UKIDSS\footnote{\url{http://www.ukidss.org}} \citep{UKIDSS-DXS}; and spectroscopic surveys
such as SDSS-III/BOSS \citep{Dawsonetal:2013}, VVDS \citep{VVDS},
VIPERS\footnote{\url{http://vipers.inaf.it/papers.html}} \citep{VIPERS2014},
GAMA\footnote{\url{http://www.gama-survey.org}} \citep{GAMA2011}, AEGIS
\citep{2007ApJ...660L...1D},
{DEEP2/3\footnote{\url{http://deep.ps.uci.edu/}}
\footnote{\url{http://deep.ps.uci.edu/deep3/home.html}} \citep{DEEP2013},
PRIMUS\footnote{\url{http://cass.ucsd.edu/~acoil/primus/Home.html}} \citep{PRIMUS2011}, }
and HectoMAP \citep{HectoMAP}.

\subsubsection{HSC Deep and UltraDeep layers}
\label{sec:deep_ud_fields}

\begin{figure*}
 \begin{center}
 \includegraphics[scale=0.45]{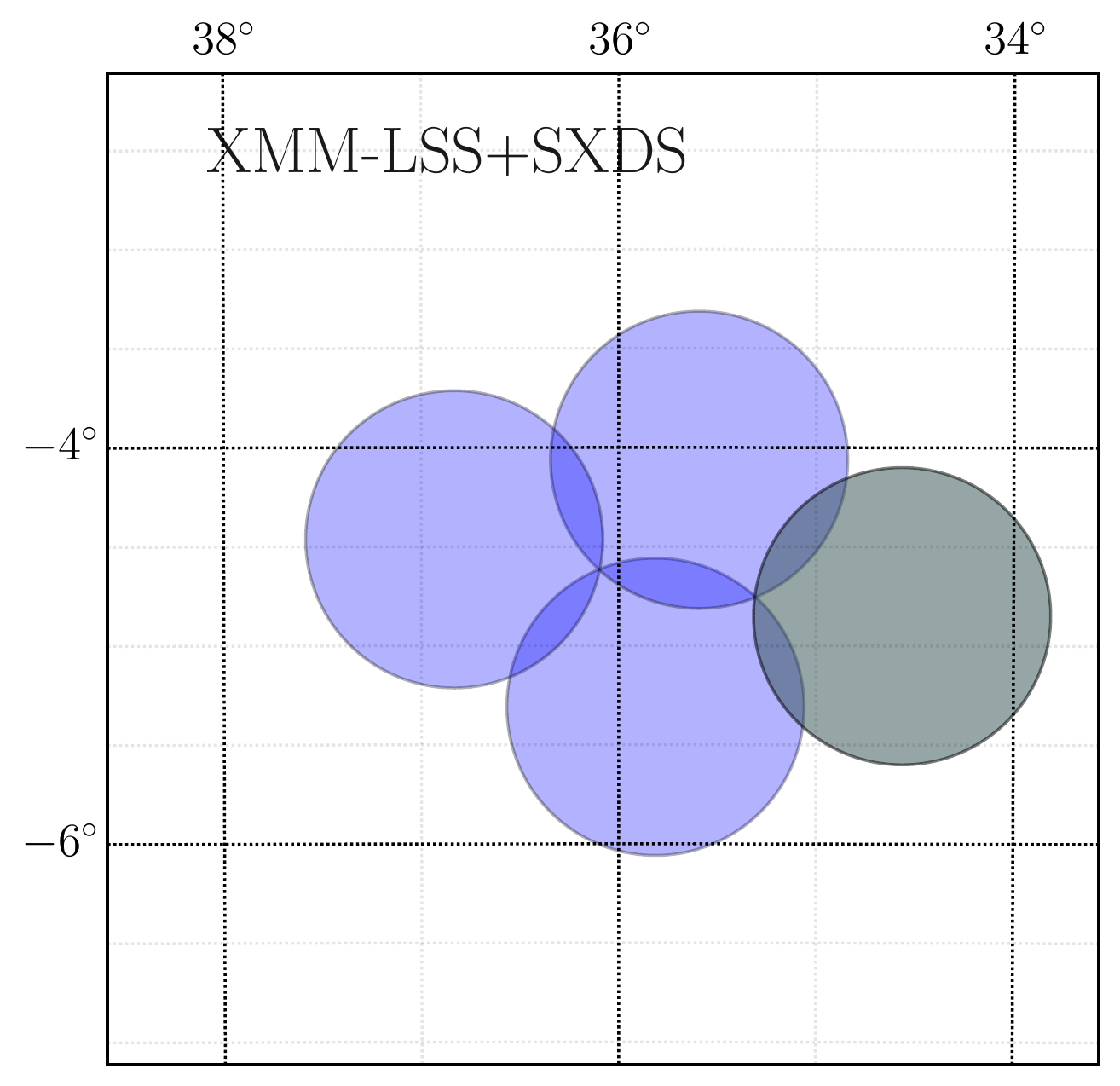}
 \includegraphics[scale=0.45]{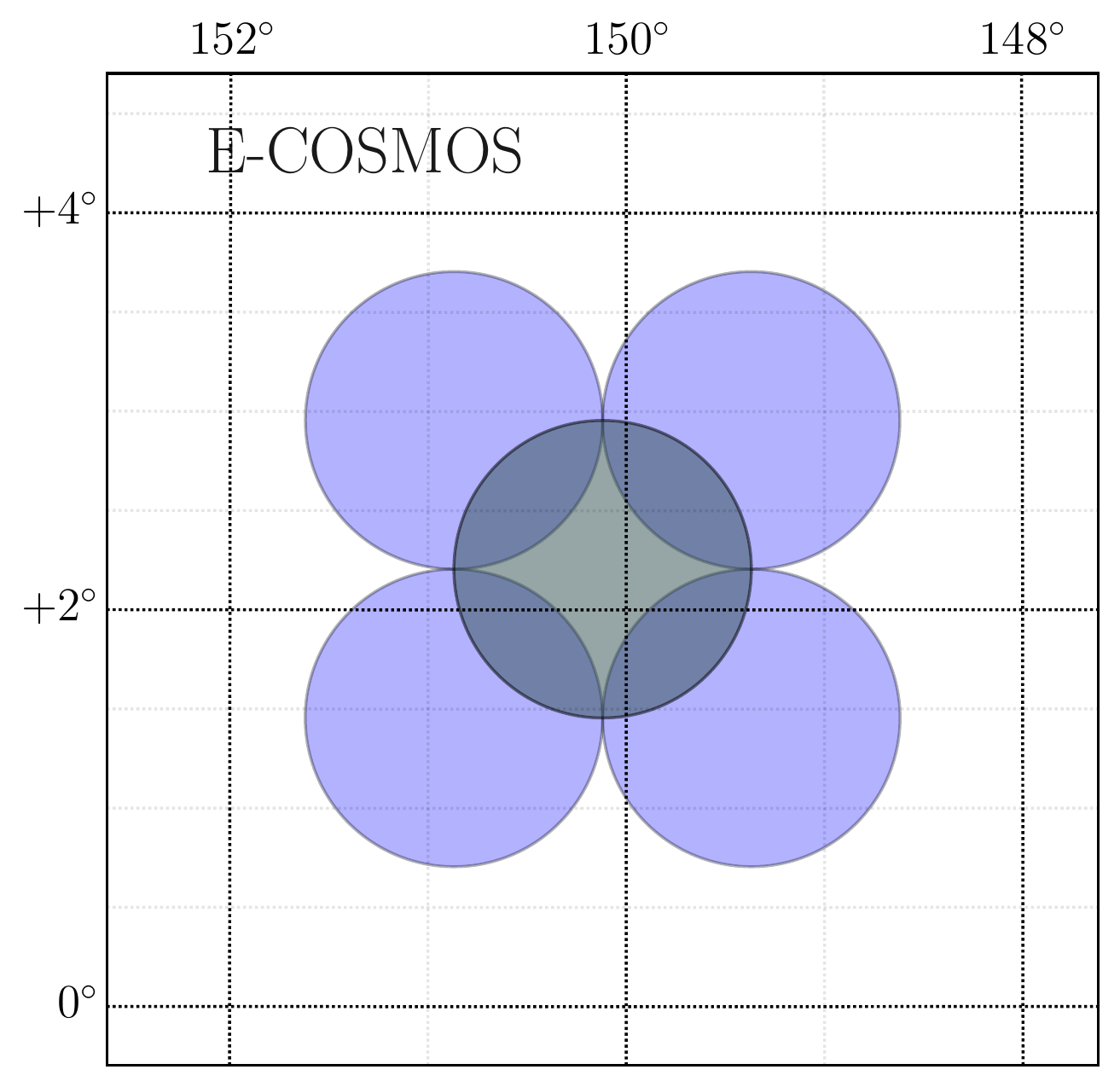}
 \includegraphics[scale=0.45]{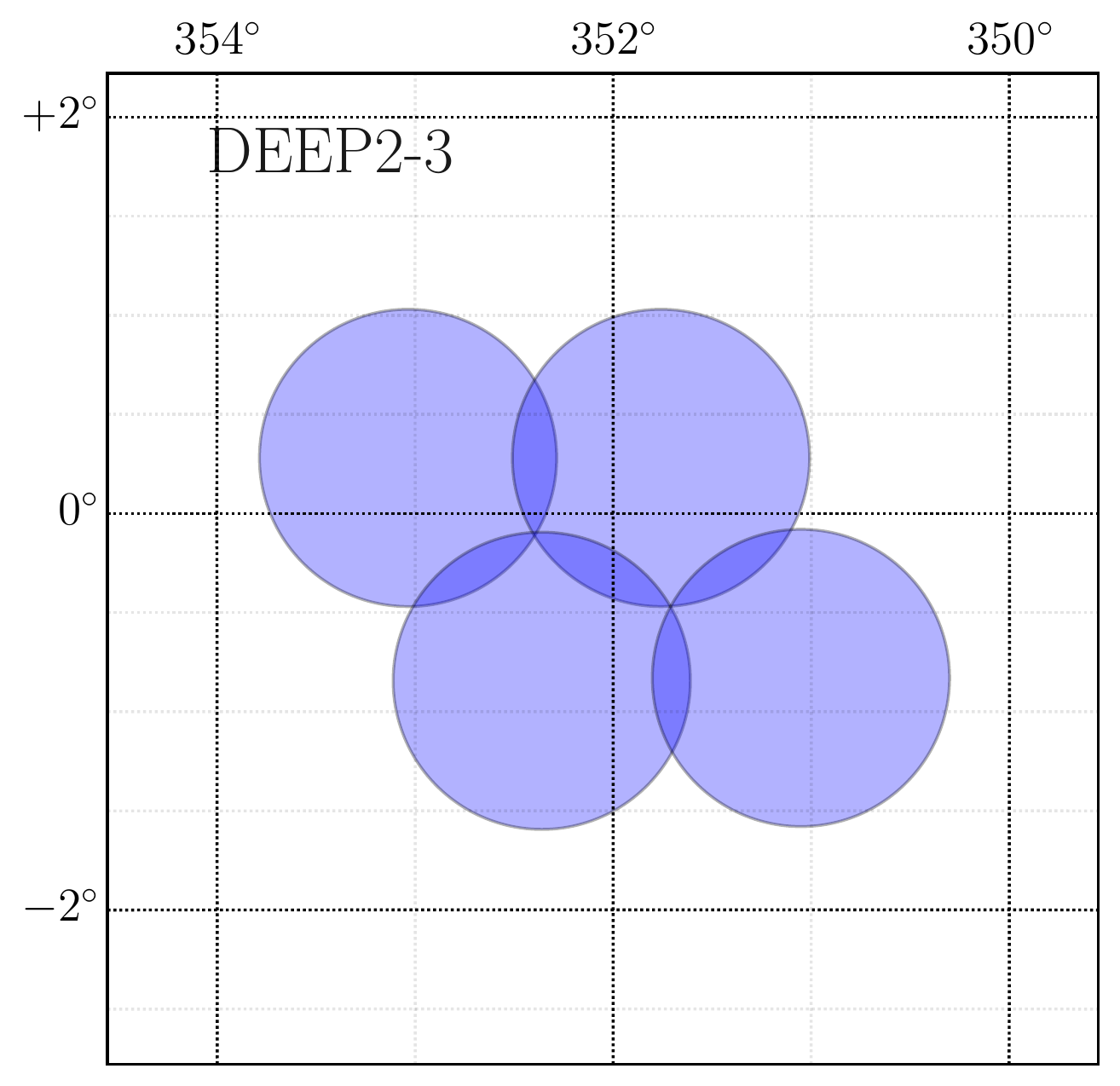}
 \includegraphics[scale=0.45]{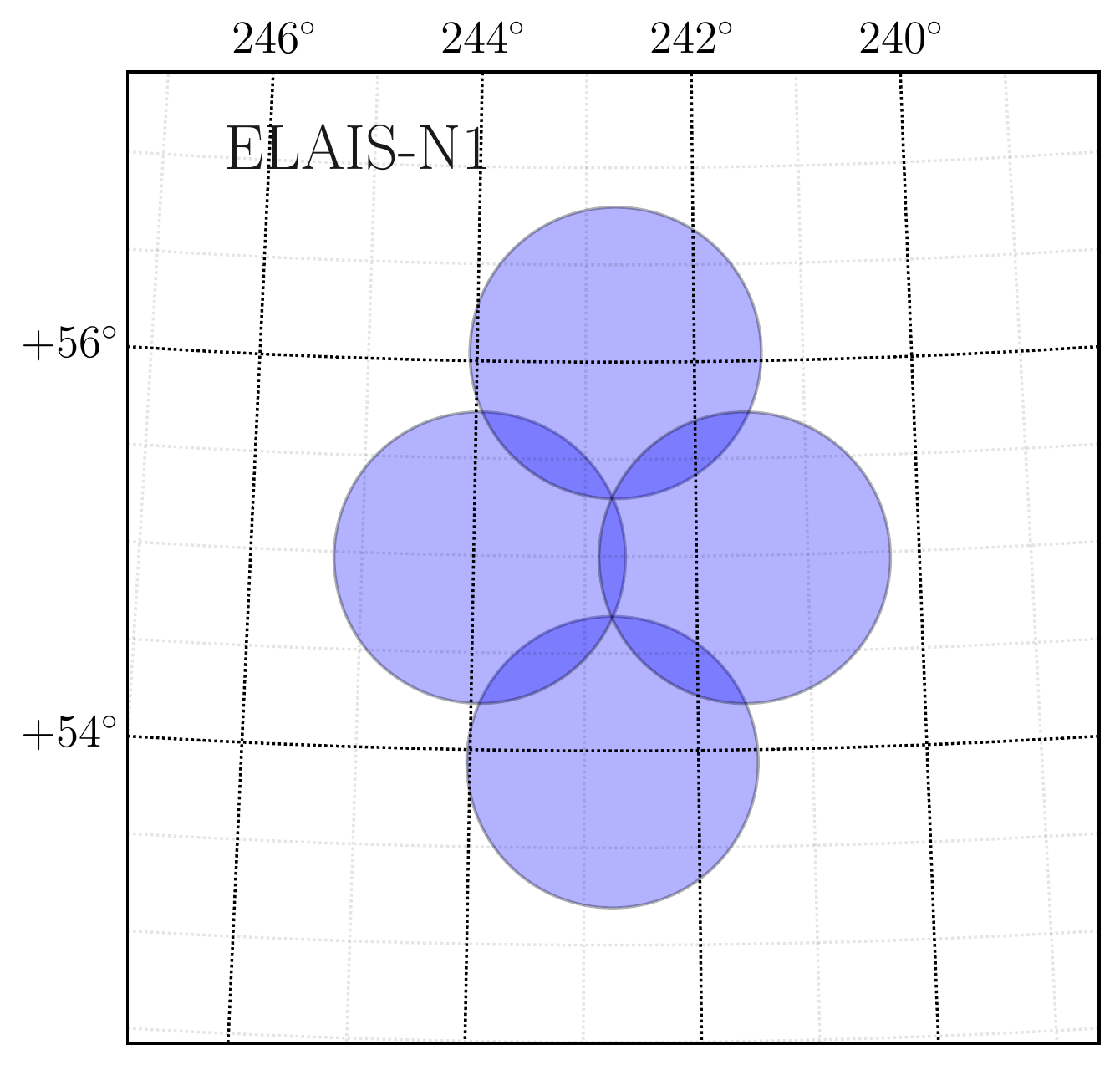}
\noindent\caption{The blue and dark-green circles show locations of the fiducial
pointings of the Deep and UltraDeep fields,
respectively (see Table~\ref{tab:field_names}). We have an additional five
dithered pointings around each fiducial pointing, as described in the text.\hfill
\label{fig:deep_fields}}
 \end{center}
\end{figure*} 
The primary science goals of the HSC-Deep and UltraDeep layers are the
study of galaxy and AGN evolution over cosmic time, and a survey
for high-redshift supernovae as a cosmological probe. The UltraDeep
regions are single pointings in the Deep fields, and (with one
exception, see below) the Deep fields are included in the Wide layer
fields (see Figure~\ref{fig:wide_fields}).  The Deep and UltraDeep
components of the survey are driven by several requirements.  The first
is to study high-redshift galaxies, including Lyman-break galaxies
selected by their broad-band colors, and Lyman-$\alpha$ emitters with
emission lines falling into the narrow-band filters corresponding to
redshifts 2.2, 5.7, 6.6, and 7.3. The second is to explore the variable
universe in the Deep and UltraDeep fields, especially searching
for $z \simgt 1$ supernovae.  At redshifts of 6 and above, the spatial
and luminosity distribution of galaxies holds important clues about the
timing and topology of cosmic reionization.  The Deep and UltraDeep data
will have significantly higher signal-to-noise ratio for galaxies at the
limits of the Wide layer imaging, making them ideal for testing
systematics in shape and photometric measurements (see
Table~\ref{tab:exptime}).

Our four Deep layer fields are listed in Table~\ref{tab:field_names} and
are shown in Figure~\ref{fig:deep_fields}.  DEEP2-3 and ELAIS-N1 each
have four pointings, while XMM-LSS has three, overlapping with a fourth
pointing to UltraDeep depths, SXDS.  There are four E-COSMOS pointings
to Deep depth, which overlaps with a fifth pointing to UltraDeep depth,
COSMOS\footnote{\url{http://cosmos.astro.caltech.edu}}.  The four fields
are all regions which already have extensive multi-wavelength imaging
and spectroscopy.  The ELAIS-N1 field does not lie in the Wide
footprint, but it has deep NIR data from the UKIDSS-DXS
\citep{UKIDSS-DXS}, and is one of the deep LOFAR \citep{LOFAR} survey
fields. LOFAR uses an array of omni-directional antennas designed to
detect the 21cm signals from neutral hydrogen in the cosmic reionization
epoch. The cross-correlation of our HSC Lyman-$\alpha$ emitter sample
with the LOFAR data will allow us to explore the relationship of
reionization to the LAE distribution.  We are gathering additional
multi-band data in the Deep and UltraDeep areas of the sky,
including $u$-band observations with the Canada-France-Hawaii Telescope
(Wang et al. in prep., Sawicki et al. in prep.), and near-infrared data
using the United Kingdom Infrared Telescope (UKIRT; Egami et al. in
prep.).  These data supplement existing VIDEO data from the VISTA
telescope \citep{VIDEO}, as well as deep pointings in the UltraDeep
fields with the Spitzer telescope \citep{Steinhardt:2014}.

In the UltraDeep layer, we will carry out the deepest HSC imaging for a
total area of 3.5~deg$^2$ in two independent blank fields well separated
on the sky, each covered with one pointing of HSC, and each overlapping
a Deep field.  Targeting two fields will yield a large sample of
high-$z$ supernovae and galaxies, and will allow us to evaluate cosmic
variance in all statistical measurements we make in the two fields.

\subsection{Observing Strategy}
\label{sec:strategy}

\begin{figure*}
\centering{
  \includegraphics[width=\linewidth]{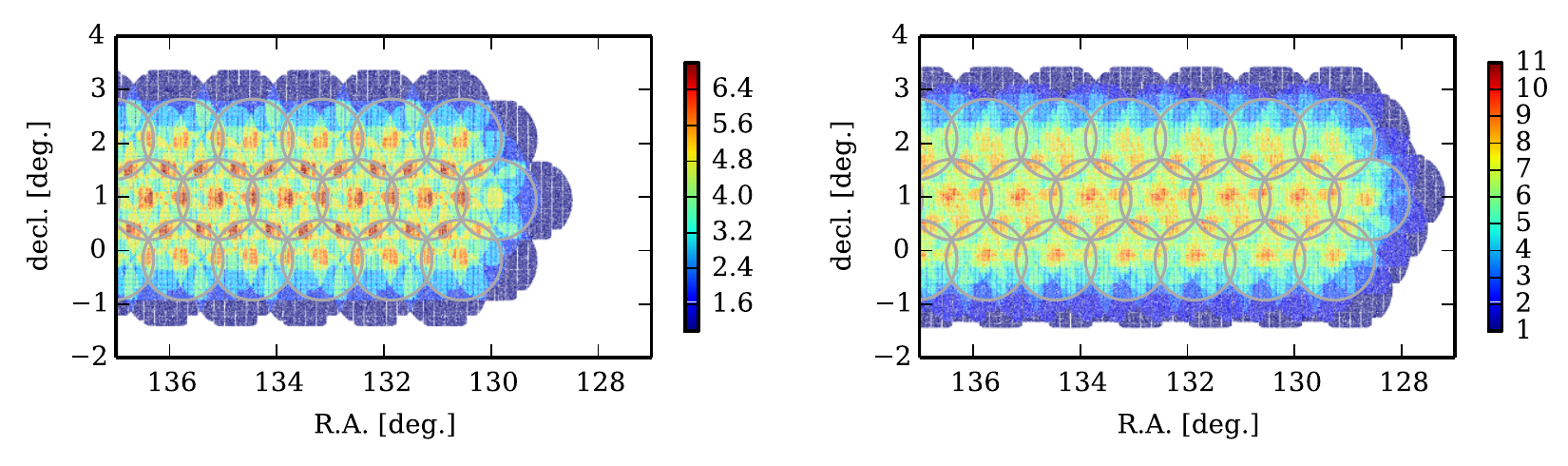}
  \caption{ The spatial distribution of the number of visits in an example
    region of the Wide layer.  The left and right panels show the
    coverage in the $g$ and $z$ bands, respectively.  The solid circles
    show the fiducial pointings around which dithering is carried out (4
    and 6 visits for $gr$ or $izy$, respectively).  Full depth
    for the Wide layer corresponds to greenish regions in each panel.
\hfill
    \label{fig:dist_countinputs} }}
\end{figure*}
\begin{figure}
\centering{
  \includegraphics[width=0.45\textwidth]{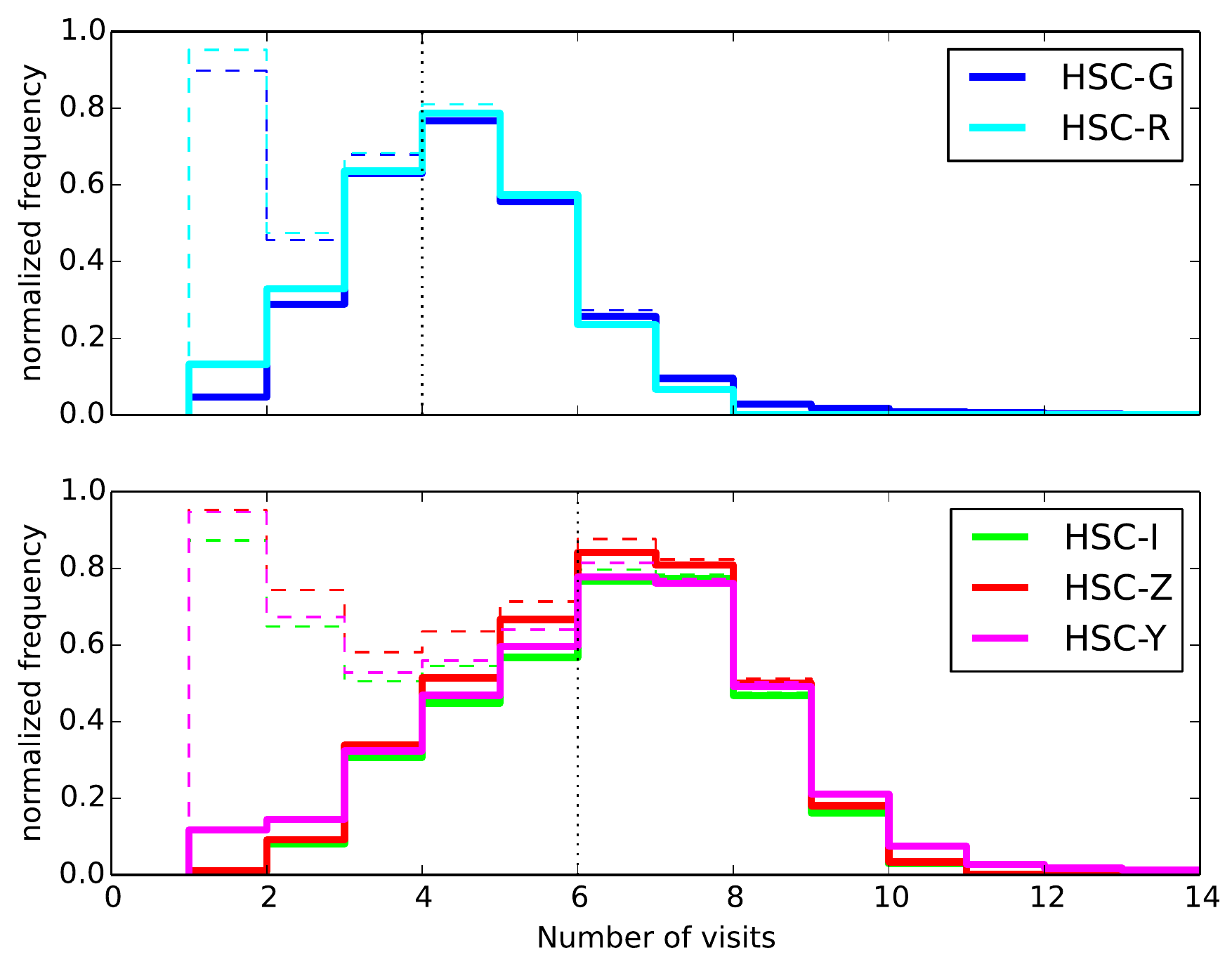}
\caption{Distribution of the number of exposures for different
    dithering patterns within the Wide layer footprint. $g$ and
    $r$-band are shown in the upper panel, and $i, z$ and $y$ in the
    lower panel, taken from the HSC DR1 data. The vertical dotted lines
    represent our target number of exposures; note that because of the
    dithering, a significant fraction of the area gets more exposures
    than this.  Dashed histograms are for the entire region, while the
    solid are for regions inside the fiducial pointings only. 
The fiducial pointings are shown in 
    Figure~\ref{fig:dist_countinputs}.\label{fig:hist_countinputs}
    \hfill}}
\end{figure}
The HSC-SSP comprises an interlocking set of observations in five
broad-band and four narrow-band filters, and with three different
layers to different depths.  Carrying out efficient observations through
the full 300 nights of the survey requires careful planning, with an
aim to making the data immediately useful to HSC scientists as the
survey progresses.  

The HSC instrument is mounted at the Prime Focus of the Subaru telescope
for each run. Runs are typically two weeks long, 
centered on New Moon.  The filter exchanger can hold six filters at a
given time, and the complement of filters cannot be changed during a
run.  Moreover, because the filter exchanger is installed the day
after the instrument is installed, and removed the day before the
instrument is removed, only one filter is available for observations on
the first and last day of any given run.  The two-week runs include
observations for the HSC-SSP, as well as a variety of general-use
programs, so the choice of filters to be installed in the exchanger for
any given run must balance the needs of all these programs.  On any
given run, the filter exchanger typically holds four or all five of the
broad-band filters, and one or two of the narrow-band
filters.

As described in the HSC DR1 paper, in the early phases
of the survey we have focused on regions of the sky with extensive
external datasets, to 
calibrate and validate our data and to maximize scientific synergy.
These included the GAMA, AEGIS, VVDS, VIPERS and DEEP2-3 regions, where
extensive spectroscopic data are available, and the COSMOS field, where
accurate 30-band photometric redshifts are also available.  In the
XMM-LSS region, we also have X-ray data from the XMM-LSS survey
\citep{XXL} and Sunyaev-Zel'dovich data from ACTPol \citep{Niemack2010},
which are complementary to optically selected \citep{Oguri2017} or
weak-lensing selected clusters \citep{Miyazakietal:2015,Oguri2017b} in HSC.  The
UKIDSS and VIKING surveys have also carried out deep near-infrared
imaging in the equatorial fields, which is particularly useful for
quasar studies \citep{Toba2015,Matsuoka2016}.

In the first years of the HSC SSP survey, we have aimed to reach full
depth in any given region of the Wide layer in all five filters fairly
quickly (i.e., within a few lunations), and only then build up area with
time.  Having photometry in all five bands is crucial for most of the
HSC-SSP survey goals.  However, given the overhead in changing filters,
we observe in no more than two filters most nights, occasionally using
three filters when observing in the Deep and UltraDeep fields. As the
survey matures, we are working to bridge already-observed fields in
order to maximize the contiguous area in the survey footprint.

If a given night is dark, and clear weather and good seeing are
forecasted, we usually start our observation in the $i$ band, in which
we will do our WL shape measurements.  We require that the $i$-band data
be taken with seeing better than $0.8''$.  If the seeing becomes
worse than this, we change filters.  The $i$-band data we have taken
mostly satisfies this condition; the median seeing in our $i$-band data
is about $0.6''$ (the HSC DR1 paper).
We perform observations in the redder filters,
$z$, $y$, NB816, NB921, and NB101 
when the Moon is up. The
seeing in other bands is only somewhat worse, and a significant
fraction of
the HSC data in all bands has seeing better than $0.8''$ (the HSC DR1 paper).

Our survey design includes about $2/3$ of the observing time in the Wide
layer, with $1/3$ for the Deep and UltraDeep observations combined.  The
narrow-band imaging in the Deep and UltraDeep fields can be carried out
only in those observing runs when the appropriate filters have been loaded
into the filter exchanger.  In the first year, we aimed to observe in
the Deep and UltraDeep layers to roughly 1/5 of the ultimate depth.
Starting in late 2016, we adopted a specific cadence for the UltraDeep
broad-band observations to maximize the sensitivity to and measurement
of the lightcurves of $z\simgt 1$
supernovae.  We are planning to obtain about 60\% of the total exposures in
$grizy$ in each of the UltraDeep COSMOS or SXDS fields 
during a single semester in 2017 or 2018 in a focused campaign to
search for such 
supernovae. This will leave about 10\% of the exposures to be carried
out over the remainder of the survey.

Finally, for each run, we usually take basic calibration data
(biases, darks and dome flats) for each of the filters installed in the
filter exchange unit.

\subsection{Pointing Strategy}
\label{sec:pointing}

\begin{figure}
\centering{
  \includegraphics[width=0.45\textwidth]{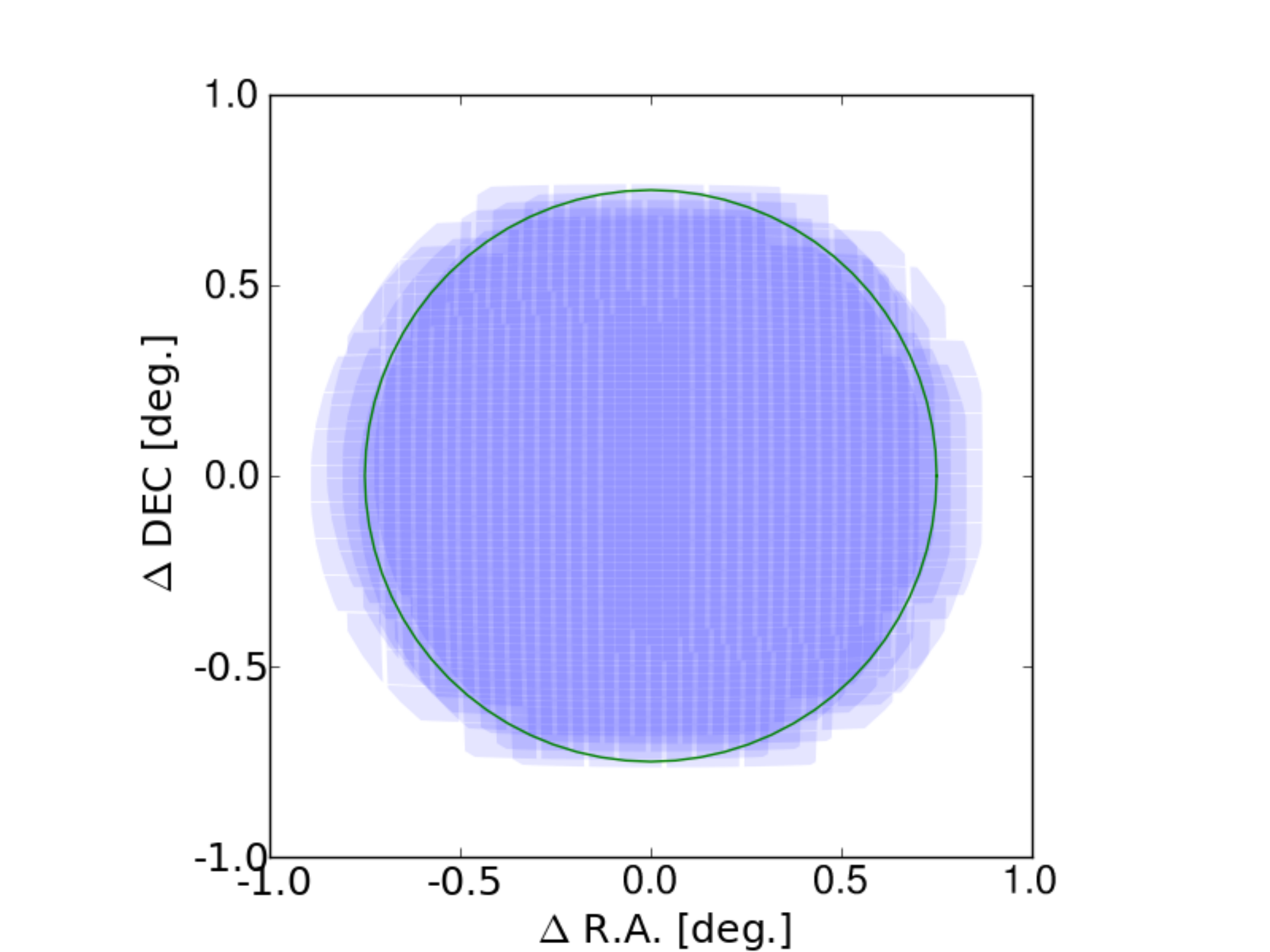}
  \caption{ The dithering pattern for one five-dithered pointing in each
    field of the Deep and UltraDeep layers.  The circle has a radius of
    $0.75^\circ$.  The intensity of the color is proportional to the
    number of visits.  The figure includes the CCD gaps.\hfill
    \label{fig:deep_dither} }
    }
\end{figure}
We have quick-look analysis tools that allow us to determine the seeing
and the sky transparency of each exposure while observing (Furusawa et
al.~2017).  The transparency is measured by comparing the observed
brightness of stars in each field with those from the SDSS.  This allows
us to make decisions on the spot regarding the filter and area of sky to
cover.  We retake visits which do not satisfy our criteria of seeing
($<1.2''$) or transparency ($>0.7$) but the exposure time is always kept
the same (see~Table~\ref{tab:exptime}).

We also monitor the focus in close to real time using data from the
focus chips, which are mounted slightly above and below the focal plane.
When they show the instrument to be out of focus, we take a special set
of short exposures over a range of focus positions, and adjust the focus
accordingly.

In the Wide layer, the total exposure time is 10-20 minutes, depending
on the filter (Table~\ref{tab:design}), divided up into 4-6 individual
visits.  We separate successive exposures in a given field by more than
half an hour, in order to have independent realizations of the
atmosphere and thereby average out atmospheric effects to some extent.
To increase the dynamic range at the bright end, we also take a single
30-second exposure in each field and for each filter.  This short
exposure gives us many unsaturated stars in the magnitude range that
SDSS and Pan-STARRS1 probe, 
crucial for the first-order astrometric and photometric calibration.

Figure~\ref{fig:dist_countinputs} demonstrates our pointing strategy to
cover the target fields in the HSC-Wide layer.  The gray circles are the
``fiducial'' pointings which define the survey geometry, each with a
radius of 0.75$^\circ$, approximately denoting the HSC field-of-view.
We dither the telescope between exposures to homogenize the depth of the
survey, fill gaps between CCD chips, improve measurement of scattered
light, and control the photometric and astrometric calibration of the
survey.  We offset the telescope between successive exposures with a
dither pattern parameterized by $(\Delta r_{\rm dith},\theta_{\rm
dith})$, where $\Delta r_{\rm dith}$ is the angular separation between
the centers of the fiducial pointing and the dithered pointings, and
$\theta_{\rm dith}$ is the position angle from the west-east direction
on the sky. We adopt $\Delta r_{\rm dith}=0.6^\circ$ for the equatorial
regions and $0.3^\circ$ for the northern sky field, which is
smaller due to the narrower width of the field.  We take $\theta_{\rm
dith}=\theta_0+(2\pi/N_{\rm dith})\times j$ for the position angle for
the $j$-th visit exposure; $j=0,1,2$ or $j=0,1,\cdots,4$ for the $gr$ or
$izy$ filters, respectively. In order to have a homogeneous depth over
different fields within the Wide layer we employed $\theta_0\simeq
63^{\circ}$ for $gr$ and $\theta_0\simeq 27^{\circ}$ for $izy$.  The
color scales in Figure~\ref{fig:dist_countinputs} show the number of
visits (exposures) at each position.

Figure~\ref{fig:hist_countinputs} shows the distribution of the number
of visits to a given region of sky for the HSC DR1 Wide layer data in
different filters.  The dashed lines are for the entire survey
footprint, while the solid lines are limited to the region within the
fiducial pointings.  Note that there are regions that go beyond the
targeted number of exposures (four exposures in $gr$, and six exposures
in $izy$).  In the DR1 catalog we define the area over which we have
full-color, full-depth data in the Wide layer to be the intersection of
the footprint in the five bands with more than a nominal number
of visits (see Section~3.9 in the DR1 paper for details).

Because the Deep and UltraDeep fields are quite small (relatively
speaking), we cannot take as large a dither as we do in the Wide layer.
We instead carry out a five-pointing dither pattern to fill CCD gaps: we
take five dithered pointings centered at ($\Delta$RA, $\Delta$Dec) $=
(0,0), (150, -150), (300, 75), (-150, 150)$, and $(-300, -75)$ (arcsec),
around the fiducial pointing given in Table~\ref{tab:field_names}. In
addition, for each set of five dithered pointings, we randomly offset
the fiducial pointing within a 7.5 arcmin radius, roughly corresponding
to a size of a CCD chip, to increase the uniformity in the field.  In
the Deep layer, the individual visit exposure times in $g$ and $r$ are
180 seconds, and 270~seconds in $i,z$ and $y$. We typically take 3-5
exposures for each field in a given filter on a given night (as long as
the weather allows).  In the UltraDeep layer, the exposure time for each
visit is 300 seconds for all bands, and we carry out 3-10 visits on a
given night.  We will continue these exposures through the lifetime of
the survey until we reach the specified total exposure times
(Table~\ref{tab:exptime}).

Figure~\ref{fig:deep_dither} shows the dithering pattern for a single
set of five exposures in the Deep or UltraDeep layer; the 
random shift is not included. The intensity of the color represents
the number of exposures covering that pointing.

\section{Data processing and HSC pipeline}
\label{sec:pipeline}

The HSC-SSP data are processed through a software package,
\texttt{hscPipe}, which is closely allied to the development of the
software pipeline for the LSST \citep{Axelrodetal:2010,Juricetal:2015}.
The spirit of the HSC image processing is that most of the core science
goals of the SSP can be carried out by the pipeline outputs.  That is,
the aim is that one will be able to do science investigations directly from
the measured 
quantities of the pipeline, without needing to re-analyze the images
themselves.  The \texttt{hscPipe} pipeline is summarized in the HSC DR1
paper, and is described in detail in \citet{Bosch2017}.  As described in
the HSC DR1 paper, the photometric calibration goal is 1\% rms
precision.  External comparison with SDSS and Pan-STARRS photometry
shows that we are
nearing that goal, but more work needs to be done.  The astrometric
calibration is good to 20 mas rms, although there are still systematic
residuals that we are working to control.

For processing purposes, the sky is divided up on a pre-defined grid
called {\tt tracts}, each covering $1.7 \times 1.7$ square degrees of
sky, and each tract is subdivided into $9 \times 9$ subareas, {\tt
patches}, squares roughly 12 arcminutes on a side.  Each exposure of the CCD
array is termed a {\tt visit}.  The data from each visit are corrected
for cosmetic features, including bad pixels and columns, and
are bias
subtracted, flat-fielded, and corrected for chip non-linearity and the
brighter-fatter effect, whereby the PSF of bright stars is larger due to
electron spreading in the CCDs \citep{Antilogus:2014}.  Then the
pipeline performs multi-visit processing to generate coadd images of
multiple exposures.  Objects are detected on the coadd images in each
band separately, and the union of these objects are used to measure the
photometric and astrometric properties across the bands.

The pipeline results depend critically on an accurate determination of
the point spread function (PSF).  This is important for the photometry
of stars, for model fits to galaxies, for determination of galaxy shapes
for weak lensing, and for modeling overlapping images (the deblending
problem).  We determine the PSF using a customized version of PSFex
\citep{Bertin2011}.  The measured properties of all detected objects are
stored in a postgreSQL database as described in Takata et al.~(2017),
while the image files are available for direct download.  Please see
\citet{Bosch2017} and the HSC DR1 paper for more details about the
pipeline, including a description of known problems with the data and
its processing.  The details of the galaxy shape catalog used for the
WL measurements are given in \citet{Mandelbaum2017}.

In addition to \texttt{hscPipe}, we also use the HSC synthetic galaxy
pipeline \texttt{SynPipe} (\citealt{Huang2017}; Murata et al. 2017). This
is a \texttt{Python}-based module that interfaces with hscPipe and 
can inject realistic synthetic stars and galaxies at desired locations
of single-\texttt{visit} HSC images. We use \texttt{SynPipe} to examine
the photometric performance of \texttt{hscPipe} \citep{Huang2017} as
well as to characterize the effects of galaxy blends for the Wide survey
(Murata et al. 2017).

\section{Conclusion}
\label{sec:conclusion}

This paper describes the design of a 300-night imaging survey of the sky
with Hyper Suprime-Cam, a 1.77 deg$^2$ imaging camera mounted on the
Prime Focus of the 8.2m Subaru telescope.  The survey will extend from
2014 through 2019, and is being done in five broad bands ($grizy$) and
four narrow bands.  The survey has three layers, termed Wide, Deep, and
UltraDeep, covering 1400 deg$^2$, 26 deg$^2$, and 3.5 deg$^2$,
respectively.  As described in detail in the HSC DR1 paper, the survey
data to date are of very high quality, with median seeing of $0.6''$ in
the $i$-band, and only somewhat worse in the other bands.  This paper
introduces a special issue of the PASJ, with a combination of
technical papers describing the instrument and survey, and science
papers describing a broad range of exciting results from the first
year of the survey.  We anticipate future data releases in 2019 and in 2021.

The HSC-SSP survey is part of a larger project, termed ``Subaru
Measurements of Images and Redshifts'' (SuMIRe).  The HSC team, together
with additional partners in the US, France, Germany, Brazil and China,
are building a wide-field multi-object spectrograph
\citep{Takadaetal:2014,TamuraPFS}, which will use the same WFC as HSC.  With it,
we plan to carry out wide-field spectroscopic surveys of stars,
galaxies, and quasars selected from the superb imaging data from the
HSC-SSP survey.

\bigskip \noindent{\bf Acknowledgements}\\ 

The Hyper Suprime-Cam (HSC) collaboration includes the astronomical
communities of Japan and Taiwan, and Princeton University. The HSC
instrumentation and software were developed by the National Astronomical
Observatory of Japan (NAOJ), the Kavli Institute for the Physics and
Mathematics of the Universe (Kavli IPMU), the University of Tokyo, the
High Energy Accelerator Research Organization (KEK), the Academia Sinica
Institute for Astronomy and Astrophysics in Taiwan (ASIAA), and
Princeton University. Funding was contributed by the FIRST program from
Japanese Cabinet Office, the Ministry of Education, Culture, Sports,
Science and Technology (MEXT), the Japan Society for the Promotion of
Science (JSPS), Japan Science and Technology Agency (JST), the Toray
Science Foundation, NAOJ, Kavli IPMU, KEK, ASIAA, and Princeton
University. HM is supported by the Jet Propulsion Laboratory, California
Institute of Technology, under a contract with the National Aeronautics
and Space Administration.  
This paper makes use of software developed for the Large Synoptic Survey
Telescope. We thank the LSST Project for making their code available as
free software at \url{http://dm.lsst.org} .

The Pan-STARRS1 Surveys (PS1) have been made possible through
contributions of the Institute for Astronomy, the University of Hawaii,
the Pan-STARRS Project Office, the Max-Planck Society and its
participating institutes, the Max Planck Institute for Astronomy,
Heidelberg and the Max Planck Institute for Extraterrestrial Physics,
Garching, The Johns Hopkins University, Durham University, the
University of Edinburgh, Queen's University Belfast, the
Harvard-Smithsonian Center for Astrophysics, the Las Cumbres Observatory
Global Telescope Network Incorporated, the National Central University
of Taiwan, the Space Telescope Science Institute, the National
Aeronautics and Space Administration under Grant No. NNX08AR22G issued
through the Planetary Science Division of the NASA Science Mission
Directorate, the National Science Foundation under Grant
No. AST-1238877, the University of Maryland, and Eotvos Lorand
University (ELTE) and the Los Alamos National Laboratory.

Based on data collected at the Subaru Telescope and retrieved
from the HSC data archive system, which is operated by Subaru Telescope
and Astronomy Data Center at National Astronomical Observatory of Japan.

This work is supported in part by JSPS KAKENHI (Grant Number
JP~15H03654) as well as MEXT Grant-in-Aid for Scientific Research on
Innovative Areas (15H05887, 15H05892, 15H05893, 15K21733).

\bibliographystyle{apj}
\bibliography{refs}

\end{document}